# Understanding Coordination in Global Software Engineering: A Mixed-Methods Study on the Use of Meetings and Slack

Viktoria Stray[a,b], Nils Brede Moe[a]
[a]Sintef Digital, [b]University of Oslo

*Abstract* — Given the relevance of coordination in the field of global software engineering, this work was carried out to further understand coordination mechanisms. Specifically, we investigated meetings and the collaboration tool Slack. We conducted a longitudinal case study using a mixed-methods approach with surveys, observations, interviews, and chat logs. Our quantitative results show that employees in global projects spend 7 hours 45 minutes per week on average in scheduled meetings and 8 hours 54 minutes in unscheduled meetings. Furthermore, distributed teams were significantly larger than co-located teams, and people working in distributed teams spent somewhat more time in meetings per day. We found that low availability of key people, absence of organizational support for unscheduled meetings and unbalanced activity from team members in meetings and on Slack were barriers for effective coordination across sites. The positive aspects of using collaboration tools in distributed teams were increased team awareness and informal communication and reduced need for e-mail. Our study emphasizes the importance of reflecting on how global software engineering teams use meetings and collaboration tools to coordinate. We provide practical advice for conducting better meetings and give suggestions for more efficient use of collaboration tools in global projects.

## 1. INTRODUCTION

Global software engineering (GSE) has become prevalent in recent years [65]. Among features that make GSE attractive are access to a larger workforce, proximity to the target market to help understand customers' needs, reduced time to market, and cycle-time acceleration [14, 57, 83]. Further, GSE often implies virtual teams, in which team members are dispersed across different locations [33]. Although the idea of globally distributed teams is attractive for many companies, it also entails some challenges. There are many obstacles that such organizations may face due to geographical, temporal, and cultural distances. For example, trust, culture, time-zone and language problems [60, 61, 63], and intra-/inter-team coordination challenges often decrease communication frequency and result in delays in communication [1, 11, 22, 43, 57].

GSE projects are seldom solved by one team alone. Coordination in a multi-team GSE environment is challenging because the work is carried out simultaneously by many development teams spread across sites [15]. Delivering results frequently and iteratively requires coordination at the project and team levels. Van de Ven et al. [95] propose three coordination modes: programming or codification (impersonal mode), coordination by feedback on the individual (personal mode), and group level (group mode). Coordination by programming in global agile projects can be exercised through Scrum guidelines, rules for quality assurance, plans, checklists, and issue trackers (e.g., Jira) [56, 62]. Examples of coordination by feedback in agile GSE projects include daily stand-up meetings, scrum-of-scrum meetings, retrospective meetings, and informal ad hoc conversations [62]. Development productivity has been found to improve when coordination needs are matched by appropriate coordinating mechanisms [12].

In the case of high uncertainty and complex projects, work relies heavily on coordination by feedback, particularly in group mode (meetings and ad hoc conversations) [95]. However, coordination by group mode can be challenging in GSE projects. Matthiesen and Bjørn [48] found that frequent interactions have little or even harmful effects when coordination practices and monitoring only provide a partial picture of the global collaboration. Therefore, GSE companies need to have a holistic view of their processes, practices, and tools. Given the complex landscape in which team members in global software projects operate, we need to understand how they coordinate by using tools and meetings and what the barriers are for such coordination.

Further, to understand coordination by meetings, we need to investigate how much time GSE project participants spend in meetings versus doing programming and testing work. While several studies claim that employees spend a large amount of time in meetings (e.g., [70, 77]), no one has investigated the actual meeting frequency and time spent in meetings in agile global software projects. Finally, distributed projects and teams need support from technology, such as electronic task boards and social software (e.g., instant messaging tools and wikis) [27]. Video systems are often used for distributed Scrum meetings [48, 92]. Teams also use collaborative instant messaging (IM) tools to mitigate the coordination challenges in distributed projects [17]. However, research on the use of such tools in globally distributed teams is scarce [2, 6, 24, 27]. In a previous paper [91], we studied the use of Slack in agile virtual teams distributed across Norway and Poland and found that one positive aspect of using the tool was increased transparency across the two sites.

To better understand successful global software development, there is a need to investigate the relationships among organizational structure, processes and coordination mechanisms [11]. Given that coordination by meetings and collaboration tools in GSE serves as essential means for group-mode coordination [17, 95], we identified the following research question:

RQ: What are the group-mode coordination challenges in GSE projects?

To investigate this research question, we studied meetings and the use of Slack in a global organization that has development sites in China, Europe, and the United States. We analyzed Slack logs from two sites, conducted 19 interviews, observed 21 meetings at four sites, and surveyed 160 people from the whole company. We also analyzed documents collected over a period of three years.

Our research contributes to the body of knowledge by providing experiences based on a longitudinal case study. The main contributions are i) an understanding of the challenges of scheduled and unscheduled meetings, ii) how much time agile project members spend in meetings, iii) an understanding of how an IM tool, such as Slack, is used to coordinate, and iv) recommendations for using Slack in GSE projects. The current paper extends our preliminary findings on the use of Slack [91].

The remainder of this paper is organized as follows: We present background in Section 2. Section 3 describes the research method used. Section 4 presents the results. In Section 5, we discuss the findings, suggest implications for practice, and discuss the limitations of the study. Section 6 concludes the paper and proposes future work.

## 2. BACKGROUND

Software development projects often involve complex activities that require multiple interdependencies among experts, roles, teams, tasks, and various software components and systems. In this section, we first present background information on coordination in a global software development context. We define coordination and introduce three coordination levels: the individual level, group level, and coordination by programming. Second, we describe different types of meetings in agile software development and GSE. Finally, we present the tool Slack and describe how it can be used as a collaboration tool for mutual adjustment in GSE projects.

### *2.1 Coordination in GSE*

Coordination has a vital role in the success of GSE [9], and informal channels of coordination play a crucial role [9]. Team productivity in GSE relies on an effective coordination structure, with both scheduled and unscheduled meetings and the right informal collaboration tools to support mutual adjustment. Often, team members communicate with each other using some form of collaboration technology [17, 38, 85]. Tell and Babar emphasize the importance of having computer-mediated teamwork tools that can support continuous coordination (delegating work to others) and show awareness (which team members are online) in globally distributed teams [64]. Further, Giuffrida and Dittrich [27] argue that the use of social software in global software development may support coordination among team members and foster awareness across sites. Although new technology and new processes have enabled better coordination in GSE [61], there are still many challenges. High pressure to master advanced communication technology and a lack of nonverbal communication due to distance, as well as problems in forming trust between distributed teams and team members, are some of the challenges [34, 37].

A widely used definition of coordination is Malone and Crowston's: "Coordination is managing dependencies between activities" [46]. This definition emphasizes dependencies, which are situational constraints on action. Sociologists Van de Ven et al. [95] define coordination as "integrating or linking together different parts of an organization to accomplish a collective set of tasks." Because Van de Ven et al. focus on the coordination of different parts of an organization (e.g., linking virtual team members and linking teams and their stakeholders), their model is highly suitable for this case study, whose focus is coordination in a distributed setting.

Coordination is performed on different levels. Van de Ven et al. [95] proposed three coordinating modes: feedback on an individual level (personal mode), the group level (group mode), and coordination by programming or codification (impersonal mode; see Figure 1). Van de Ven et al. [95] argue that coordination (linking together different parts or the organization) happens through communication channels or communication mechanisms. Once implemented, the impersonal coordination mechanisms are codified and require minimal verbal communication between people. In the personal mode, individuals serve as the mechanism for making mutual task adjustments through either vertical or horizontal channels of communication. Group mode involves a group of individuals, and typically include group meetings, whether scheduled or unscheduled. This latter division is intended mainly to differentiate between the more routine encounters and informal conversations between coworkers [95].

*2.2 Meetings in agile software development*

While regular and scheduled interaction is vital in distributed teams [60], research also shows that developers need ad hoc and informal communication [31, 40]. In agile software development, group mode coordination by scheduled meetings at the team level is ensured through practices such as planning meetings, daily meetings, demonstrations, and retrospective meetings. In Scrum, a self-managing team develops software in increments (sprints); each sprint starts with a planning meeting and ends with a retrospective and a review meeting. The team coordinates on a daily basis through a 15-minute daily meeting, which is the most common agile practice adopted by 86–87% of agile practitioners [90, 96].

Despite being a popular practice, the daily stand-up meeting is difficult to conduct successfully [92], and studies in global software development have found the meeting to be disengaging, repetitive, and boring

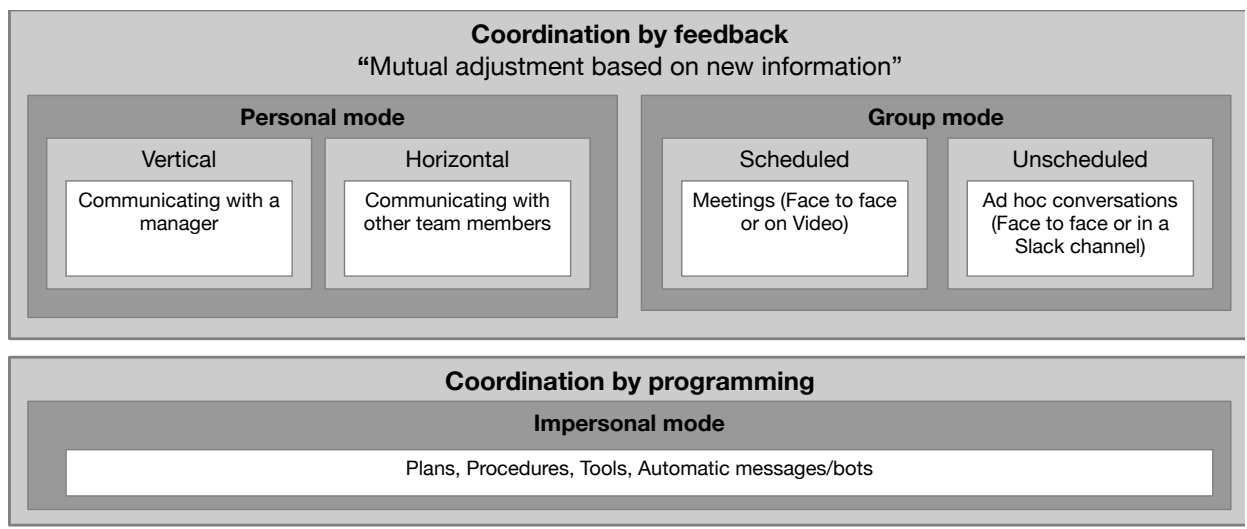

**Figure 1.** Overview of coordination modes (based on Van de Ven et al. [95])

[44]. The same has been found for scrum-of-scrum meetings (stand-ups for coordinating multiple teams) [66]. In agile projects, features or requirements to be implemented are registered in a product backlog. Multiple stakeholders such as architects, domain experts, clients, project teams, designers, marketing and salespeople, management, and support can participate in the planning phase (usually through meetings) to identify the product backlog items. During the planning meeting, the product owner is responsible for presenting a prioritized product backlog to the team. The highest priority items from the product backlog are then detailed in a sprint backlog during a team-planning meeting. Because the team and the product owner are responsible for defining and improving coordination practices, what meetings are conducted and how they are conducted will change over time.

Group-mode coordination via unscheduled meetings is best ensured at the team and inter-team level by team members and teams sitting together in the same office. Nyrud and Stray [62] observed that informal and ad hoc conversations emerged in a large-scale web-program as a result of teams being co-located in an open office. However, in distributed projects, co-location is not possible, reducing the level of ad hoc conversations. Although a distributed agile project needs to work on how to enable unscheduled meetings, and many scheduled meetings and forums increase the amount and frequency of interactions between teams, Šmite et al. [29] found it was difficult to hold unscheduled meetings in a distributed program because of too many scheduled meetings.

Relying on meetings for coordination is challenging in a distributed organization. In distributed software projects, coordination by group mode can take part within teams, between teams and managers, or among groups of team representatives acting on behalf of their teams. In a distributed setting, a common strategy for coordination across sites is to have one team member act as "ambassador" to participate in meetings with ambassadors from other teams, and management. Research on meetings is important because it might significantly reduce associated costs.

Meetings have direct costs in the form of salaries and participants' time, as well as indirect costs such as time away from primary tasks, employee stress, and reduced job satisfaction [71]. We use the following definition of a meeting: "A meeting is a gathering of two or more people for purposes of interaction and focused communication" [97]. Some of the primary purposes of meetings are to share information, solve problems, make decisions, develop strategies, or debrief teams [58]. A study of how software developers spend their time found that developers spend 1 hour in planned meetings and 0.5 hours in ad hoc meetings per day [52].

*2.3 Slack as a collaboration tool for mutual adjustment*

A lack of awareness of what distributed members are doing raises the barrier for initiating contact [30]. Furthermore, receiving delayed feedback is a challenge often faced by distributed teams, and the time to receive a response increases dramatically with the use of only asynchronous collaboration tools [32]. Casey and Richardson [10] studied two distributed teams located in Ireland and Malaysia that used e-mail and found that use of only this type of asynchronous communication also increased the misunderstandings and ambiguity of information. Instant messaging (IM) tools support team coordination and reduce the challenges of geographical distance by letting team members and managers in global projects take part in daily communication to know what difficulties the team is facing at any time [26, 55]. Calefato et al. [7] found that text-based communication may be preferred over face-to-face communication for requirements elicitation in GSE teams and when discussing openly conflicting issues. Further, Dittrich and Giuffrida [17] found that IM tools enabled awareness, were perceived as less intrusive than phone calls and enabled informal and almost synchronous communication across sites. However, Matthiesen et al. [48] found that it was essential that the people communicating on IM in GSE knew each other and had good relationships, in order for the inquiries not to be perceived as disruptive.

Slack is an IM collaboration tool launched in 2014 that now has 12 million active users daily [80]. In addition to IM, it is also possible for users of Slack to make voice and video calls and share files. Slack supports coordination through feedback on the group and personal modes and both vertical and horizontal channels. Williams [98] described how Slack can be seen as a chat room, where the whole company and its different teams can be broken into smaller channels for group discussion. These channels can either be public or private. Public channels are visible to the entire team, and all the team members can join them, while private channels require an invitation to join. Slack was built around the principle of having easy access to information; its name is an acronym for the phrase "searchable log of all conversations and knowledge" [81]. Slack is multiplatform software, meaning it can be used on all kinds of operating systems, including mobile phones.

Messages in Slack can be seen and searched for only by those who are involved in the messages. Using emojis is a way of communicating in Slack and can be used when composing a message to communicate emotions. Readers of a message can react to it and communicate their emotions to the sender using emojis as well. It is also common to get the attention of some or all the members of a channel by using the "@" symbol. In this case, users receive a notification immediately if they are online. When communicating on Slack, users can edit their posts without any time limitation, and "(edited)" will be shown after the message so that the others know that the message is changed. As a consequence, a person may respond quicker than and without as much thought as when writing an e-mail, knowing he or she can edit the message at a later time. A person may also delete the message if they regret sending it. These features may lower the threshold to communicate with others in a distributed project.

As Slack is quite a new tool, there is little research on the use of this specific IM tool yet. However, Stoeckli [85] looked into Slack chatbots and integrations from an affordance perspective and explored their constraints within enterprises, while Lin et al. [42], through an exploratory study, tried to discover how developers use Slack and what kind of benefits it could give teams. In a recent study of agile distributed software development by Lous et al. [45], they found that the adoption of Slack eliminated the use of e-mails for internal communication. Moreover, Calefato and Lanubile [8] proposed a model to integrate developer tools with Slack as a hub. In an earlier study of a large-scale agile project, we found that Slack managed knowledge, processes, and resource dependencies [89]. Coordination on Slack can be categorized as vertical when team members communicate with their manager and horizontal when communicating with their team members. We argue that when project members communicate in group channels, they are exercising group-mode coordination because a group of people can follow and join the conversation.

## 3. RESEARCH DESIGN AND DATA COLLECTION

To answer our research question and to investigate coordination in GSE, we used a longitudinal single-case holistic study [101]. We chose to perform a longitudinal case study because previous studies on coordination in software development show that coordination changes over time [16]. In addition, we believed that having a deeper understanding of a single company over time would give us better opportunities to collect quantitative data to examine the use of meetings in the whole company. Šmite et al. [83] argue that choice of coordination mechanisms depends on context, so studies on team coordination in global settings must make sure to describe the context in detail. We describe the context in the following section.

### 3.1 Case investigation contexts

The company we chose to study, called Geosoft (a pseudonym), is a large software company that produces and sells specialized software for the engineering domain. Geosoft develops both mass-market software and customer-specific software on a contract basis. The company has more than 15 years of experience with global software development. In addition to developing software in its main offices in Norway, the company also develops software in its offices in Poland, Germany, China, Malaysia, the United States, and the United Kingdom.

**Table 1**
Overview of data collected.

|  | **Poland–Norway** | **China–Norway** |
|---|---|---|
| Teams | Local and distributed | Local and distributed |
| Number of semi-structured interviews (avg length in minutes) | 8 (56 min) | 11 (64 min) |
| Roles interviewed (number of people) | Testers (5)<br>Developers (3)<br>Tech leaders (2)<br>Managers (2) | Testers (10)<br>Developers (1)<br>Team leaders (4)<br>Managers (5) |
| Scheduled meetings observed (Number of meetings) | Stand-up meetings (4)<br>Retrospective meetings (4)<br>Demo meetings (1)<br>Scrum-of-Scrum meetings (1)<br>Seminars (1)<br>Workshops (1)<br>Task force meetings (1)<br>Bug triage meetings (1) | Stand-up meetings (2)<br>Retrospective meetings (3)<br>Workshops (2) |
| Slack logs | Yes | No |
| Survey | Yes | Yes |
| Documents | Roadmaps, organizational charts, method description, role descriptions, project descriptions, quality assurance processes, process descriptions, retrospective documents, Wiki pages and audit reports | Organizational charts, method description, role descriptions, Team descriptions, project descriptions, quality assurance processes, process descriptions, retrospective documents, Wiki pages and audit reports |

Motivated by the need for more studies on globally distributed software teams, and to understand approaches to and solutions for coordination in such teams, we studied developers, testers, team leaders, tech leaders, and managers in the global company. We studied more than 50 members at the sites in Norway, Poland, and China.

We have collaborated with the company since 2010. Our first investigations in this longitudinal study concerning coordination took place in 2015. Because coordination is affected by time and cultural distance [1], we chose to study one nearshore project and one far-shore project. We collected qualitative data from six teams; four of the teams were distributed between Norway and Poland, and two of the teams were distributed between Norway and China. The nearshore project (the Norway–Poland case) was chosen because the project members had collaborated for a long time and the project was one of the first in Geosoft to use Slack as a collaboration tool. Furthermore, the project had no time difference and participants were close culturally. The far-shore project, the China–Norway case, was chosen because it represented the project that had the fewest overlapping hours—three hours' overlap in the summertime and only two hours in the wintertime. This makes coordination and the use of meetings even more challenging. We interviewed and observed the team members in all countries, and we also collected and analyzed Slack logs, meeting minutes, and other project materials.

*3.2 Data collection*

We had six main data collection rounds, as shown in Figure 2. Throughout the study, data collection and analysis occurred within an iterative process. Also, we worked iteratively, alternating between analyzing data and collecting new data from the nearshore project and the far-shore project. We decided to use a mixed-methods approach to investigate coordination mechanisms using qualitative and quantitative data. We distributed a survey to participants in Geosoft to gain a better understanding of how employees were coordinating. Table 1 shows an overview of our data collection.

*3.2.1 Interviews and observations*

We observed and interviewed team members in four Geosoft locations in three countries. We conducted interviews and observed scheduled and unscheduled meetings in Norway between March 2015 and August 2017, in China in April 2016 and 2017, and in Poland in September 2017 (see Figure 2). We also had several informal conversations and unstructured interviews with the project members from all sites throughout the study.

We conducted 19 semi-structured interviews; eight interviews in the Poland–Norway case and 11 interviews in the China–Norway case. More than half of the semi-structured interviews were group interviews, and the total number of interviewees was 32 people (20 females and 12 males). An overview of the roles of the interviewees can be found in Table 1. Five different roles were interviewed in total. If interviewees had more than one role, the table shows their main role in the team. The participants gave their consent for the interviews to be recorded and agreed to the publication of the results subject to anonymity. The interviews varied between 40 and 102 min. Both authors participated in most interviews. One asked questions, and one took notes and asked additional questions at the end.

Each interview consisted of four parts. In the first part, we introduced ourselves and assured interviewees of confidentiality. The topic of investigation presented to the interviewees was "coordination and teamwork in global software development." The second part comprised questions regarding the interviewee's background, experience, and current activities. The third part involved the main interview and included questions about coordination, communication, Slack, meetings, and teamwork in general. The third part was modified during our longitudinal study based on iterative data analysis. The fourth part included closing questions, and we provided an opportunity for interviewees to ask questions and make additional comments. Appendix A shows the interview guide as of April 2017.

We observed a total of 21 scheduled meetings. The majority of these meetings were virtual, involving participants from two or more sites. Our participant observation was guided by a protocol based on Spradley [84], which contained questions to be answered by the researcher and stayed constant throughout the study (Appendix B). Information recorded while observing meetings included names and roles of attendees; start and end times; types of discussions, leadership, and facilitation; format (who was sitting or standing); type of technology used (e.g., phone or video); and number of participants. In three of the meetings, we drew conversation maps to see the flow of the conversations. That is, if one person talked to another person in the meeting, we drew a line between those two persons.

*3.2.2 Survey*

Based on analysis of the first interviews we conducted, the results suggested that meetings, and especially daily stand-up meetings, were important for coordination across sites. Therefore, to better understand group-mode coordination, we conducted a survey (see Appendix C) within Geosoft in the summer of 2017. We received 160 responses across nine countries, and have previously analyzed a subset of this survey (66 responses), reported in [86]. The survey was administered through Qualtrics software. We used a five-point scale on all Likert questions. As a basis for the survey, we used the same questions and survey scales as a study of programmers, reported in [90]. We presented nominal-scale questions in a randomized order of categories because the order of response alternatives can influence results [79]. Not all questions were compulsory, which resulted in gaps in data for some variables.

Most respondents received the surveys from their leaders via e-mail. The respondents completed the survey in eight minutes on average. We received 160 responses, of which eight had to be removed because they were incomplete. Of the 152 analyzed respondents, 32 were female and 106 male, and 14 did not specify gender. The mean age of the respondents was 38.9 years, and they worked in Poland (13), Norway (36), China (17), Belgium (2), Germany (19), Hungary (1), Singapore (5), the United Kingdom (27), and the United States (32). We received responses from 71 developers, 34 testers, 32 managers, and 9 software architects. Six people reported other roles, such as designers.

*3.2.3 Slack logs*

We chose to study the use of Slack by four of the agile virtual teams (30 project members in total) in a Geosoft product center. We chose a product center distributed across Norway and Poland because its staff were mature in agile methods and had used Slack as a collaboration tool since 2015. The teams co-located several times a year and experimented with a variety of tools and processes. At the time of the interviews, the four teams were named front-end, back-end, operations, and user experience.

We collected Slack logs that included approximately 30,000 messages sent between team members in Norway and Poland over 2.5 years, from March 2015 to August 2017 (Figure 2). The messages were sent across 70 different Slack channels. Most channels were no longer active and had been archived, indicating that Geosoft adapted Slack to current needs. We could not access direct messages between two individuals because of privacy settings in Slack. We did not include Slack logs from the China-Norway collaboration since less than half of the people involved used Slack at the time of the study. Although we did not analyze Slack logs in the China–Norway case, we asked about Slack in interviews and meetings.

Chat logs exported from Slack were in the form of a compressed zip file that contained several folders, which represented the channels in Slack. Each folder included many JSON files, with each one containing the chat logs for one day. Inside the JSON file, every message was coded in a special format showing the ID of the sender, the message, and a timestamp. We had to convert the IDs to usernames, and the timestamps, which were in Epoch format, to a readable format. Finally, it was not possible to import JSON files directly into NVivo, so we converted the files to PDFs before importing them into NVivo.

*3.3 Data analysis*

All interview transcripts, observational notes, documents, and Slack logs were imported into NVivo, coded, and discussed among the authors. We also imported a Qualtrics report of the survey results for a qualitative initial coding of the results. For a detailed analysis of the user activity in the Slack logs, we used Excel; and for quantitative analysis of the survey, we used R statistical software.

When analyzing the survey in R, we looked for data in NVivo (the interviews, observations, documents, and Slack logs) that could explain our findings. For example, we found that 84% of respondents attended the daily meetings, and that 20% of the respondents attended both local and distributed daily meetings. From our qualitative data, we knew that a project team in Norway had six different daily meetings scheduled with their Polish colleagues, and that some felt they could not start working until after lunch because of all the daily meetings in the morning.

Throughout the research, we wrote memos that acted as a log and gave us the ability to look at how our reflections evolved and why particular decisions were made during the research. Writing memos is a technique inspired by grounded theory research to record reflections on the data and codes and their relationships as they occur to the analyst while coding and writing [28]. Our memos usually consisted of a few statements or questions. An example of a memo written during the analysis had the title "many participants" and contained the question, "Does the number of participants affect meeting satisfaction?"

In NVivo, we coded parts of the documents manually and applied descriptive coding. Descriptive coding uses a word or a short phrase to summarize the topic (as opposed to the content) and is useful for answering questions such as "What is going on here?" [75]. During the analysis, we assigned pieces of text to a descriptive code ("node" in NVivo). These codes were grouped and categorized into categories we believed were significant to understand group mode coordination. See an example of our coding in Figure 3. From the descriptive codes, eight different categories emerged.

We also identified a list of scheduled and unscheduled meetings, which made it possible to identify which meetings were involved when understanding a category. We coded the following meeting types:
- Agile meetings
    - Stand-up meetings
    - Planning meetings
    - Retrospective meetings
    - Demo meetings
    - Scrum-of-scrum meetings
- Other meetings
    - Weekly meetings
    - Seminars
    - Communities of practice

- o Task force meetings
- o Bug triage meetings
- o Handover meetings
- o Unscheduled meetings
- o Water cooler meetings
- o Chat meetings
- o Workshops

When we identified a phenomenon related to a specific meeting in the interviews, we searched for information about the same type of meeting in the observations and in the Slack logs. One example was the retrospective meeting, for which we found discussions on how to conduct and facilitate retrospectives in the Slack logs.

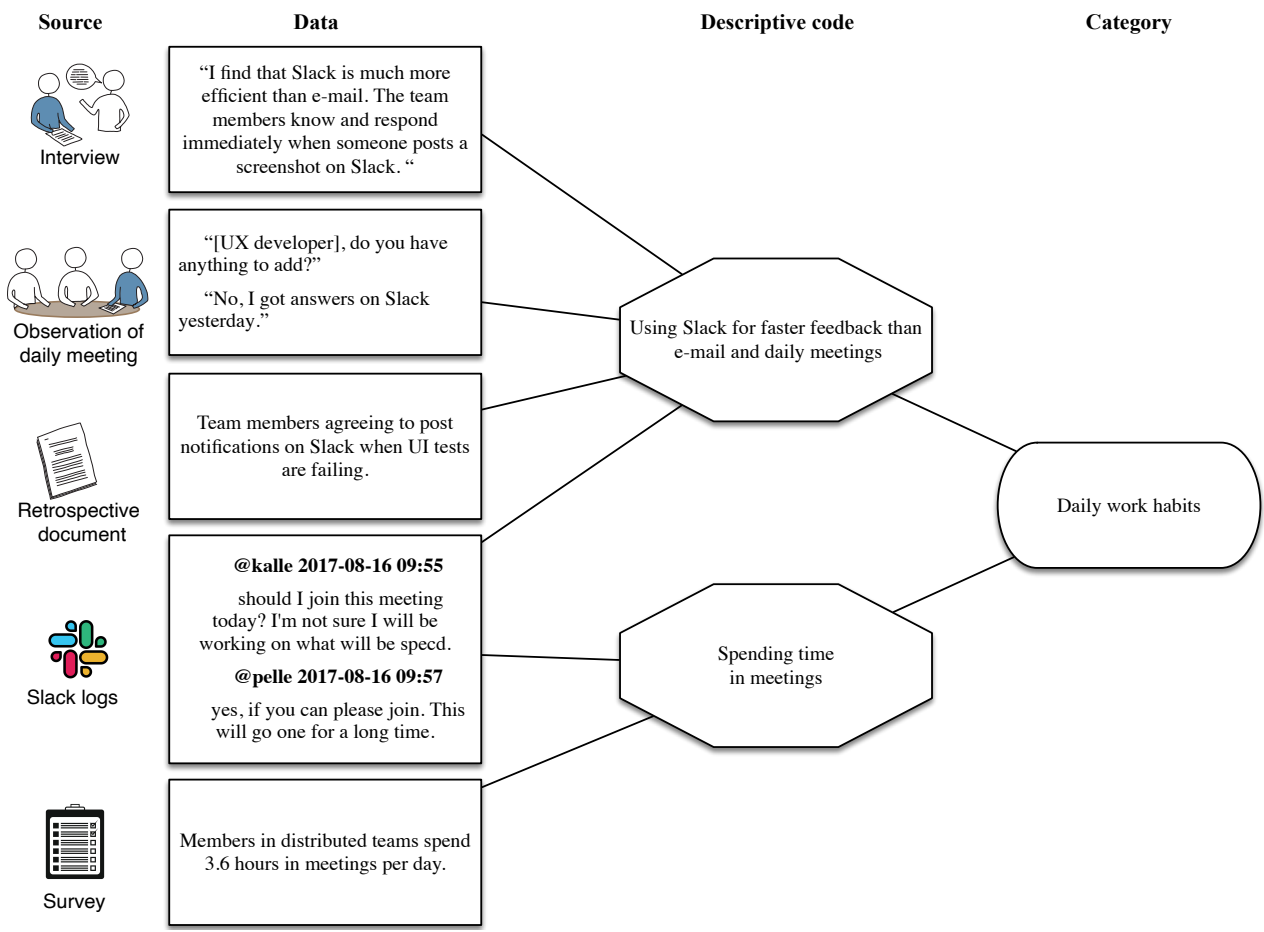

**Figure 3.** Example of data sources and analysis

## 4. FINDINGS

In this section, we first present the background and context of the studied case, such as the agile method used and team sizes. The agile GSE context is important to understand the phenomena being reported. We then present the observed phenomena that we have grouped into eight categories related to group mode coordination: 1) daily work habits, 2) availability of key people, 3) unbalanced activity, 4) meeting satisfaction, 5) co-location, 6) Slack collaboration, 7) unscheduled meetings, and 8) awareness.

## 4.1 Agile GSE context

Geosoft has projects that are either allocated in one site or shared between two or more locations. One team member or manager typically participates in more than one project. We found that many employees reported being part of a globally distributed team as well as a local team, which is natural given the project structure for the globally distributed company. For example, many attended both local and distributed stand-up meetings. In the survey, we asked, "Consider the team that you spend the most time in. Is your team distributed or co-located?" Fifty-two percent responded that they work in co-located teams, and 48% stated that they work in a team with members distributed across sites. The interviews with the Chinese participants confirmed that they felt they had a local team, but were also part of a bigger distributed team.

The average team size was seven members (see descriptive statistics in Table 2). When analyzing the difference in team size for local versus distributed teams, we found that the distributed teams were significantly larger than co-located teams ($p < 0.001$). The mean number of team members in co-located teams was 5.8, while distributed teams had an average of 8.7 members.

The company introduced agile methods in 2008, and most projects currently relied on agile methods (93%). The participants in the survey stated that they used Scrum (51%), Kanban (13%), or ScrumBan (29%; a combination of Scrum and Kanban); only 7% stated they still used Waterfall. The mean length of their sprints were 3.3 weeks. Further, the teams reported using agile practices such as planning meetings, daily stand-up meetings, retrospective meetings, demos, grooming meetings, and group learning meetings. Additionally, regular face-to-face meetings were essential for running their distributed projects and improving the agile practices. The team members were seated in an open work area.

While the organization wished to standardize collaboration tools as much as possible in order to enable everyone to communicate with everyone else in the same company (changing projects should not require a change of tools), they also let projects and teams use technology that suited their specific purposes. Often, when we visited the company, the employees were using new technology or a combination of technology. "Many of the team members appreciated having frequent video meetings, combined with using Slack." As one Chinese tester stated: "The current way with Slack and videoconferencing is the way our team likes it."

People generally prefer to communicate in their native languages, even though the official working language is English (as was the case in Geosoft), because using one's native language is more comfortable and faster. As one team leader in China stated, "Some developers on my team are Chinese, so we can talk to each other in Chinese. It is more efficient than speaking English." The teams in our study acknowledged the need to use English in Slack and discussed this at a retrospective meeting. However, we found that team members still wrote messages in their native languages when communication was directed to people in their own countries. Further, in the Polish case, some developers were not very proficient in English, which probably reduced their capability and motivation to write in English. As one Norwegian tech lead explained, "We have struggled somewhat because of language issues. Domain knowledge in combination with low language skills has made it difficult. It was a real problem that one person in particular was really bad at English writing." The use of native languages in Slack prevented team members at other sites from understanding and engaging in parts of the communication.

**Table 2**
Descriptive statistics.

|  | Unit | Respondents | **Mean (M)** | s.d. | median |
|---|---|---|---|---|---|
| Meetings | Frequency per day | 139 | **2.29** | 2.0 | 2 |
| Time in scheduled meetings | Hours per day | 144 | **1.55** | 1.8 | 1 |
| Time in unscheduled meetings | Hours per day | 144 | **1.78** | 1.7 | 1 |
| Total time in meetings | Hours per day | 144 | **3.32** | 3.3 | 2 |
| Time programming/testing | Hours per day | 144 | **5.88** | 3.3 | 6 |
| Team size | Members including self | 141 | **7.19** | 3.5 | 7 |

*4.2 Daily work habits*

From the survey, we found that those in both distributed and co-located teams attended the same number of meetings per day, which was two meetings per day on average (managers attended 3.5 meetings per day; developers, 1.6; testers, 2.1; and architects, 2.3). However, those in distributed teams spent somewhat more time in meetings per day (co-located, 3.0 hours; distributed, 3.6 hours). Overall, the employees spent 16 hours and 36 minutes per week in meetings. Looking at the time spent in meetings per site for the major sites in the company, the participants in Norway spent 3 hours in meetings per day on average; in China, 2.1 hours; Germany, 2.5 hours; the United Kingdom, 3.4 hours; and the United States, 4 hours. In Poland, employees spent 2.5 hours per day in meetings (most survey respondents from the Polish site were developers, and, few managers from the Polish site responded to the survey). Developers and testers spent the same amount of time in meetings across sites, which is natural, as they follow the same type of process (agile methods).

On average, the respondents spent 1 hour 33 minutes per day in scheduled meetings and 1 hour 47 minutes in unscheduled meetings and ad hoc conversations (see Table 2). All roles except managers said that they spent more time in unplanned coordination (unscheduled meetings and ad hoc conversations) than they did in planned coordination (scheduled meetings). A team member typically used between 6 and 7 hours on programming or testing. Specifically, the developers spent 5.4 hours per day programming and 1.6 hours testing, while the testers spent 5.7 hours testing and 0.5 hours programming.

We found that there was a relationship between a person's role and his or her time spent in meetings. As Table 3 shows, the developers and testers spent approximately 2 hours per day in meetings (both scheduled and unscheduled). During a typical work week, they spent 7 hours and 45 minutes in scheduled meetings. The managers reported spending as much as 14 hours and 21 minutes in scheduled meetings and 12 hours and 42 minutes in unscheduled meetings per week. That is, they spent more than twice as much time in meetings than developers (Table 3).

The stand-up meeting was the most common practice for coordination across the distributed team members, and involved most of the roles in the organizations; therefore, we investigated this type of meeting in detail. As many as 84% of those working in Geosoft attended daily stand-up meetings. Specifically, 40% of the respondents attended local stand-up meetings, 24% attended distributed stand-up meetings, and 20% attended both local and distributed stand-up meetings. Some sites organized several different stand-up meetings every day. For example, a project team in Norway had six different daily meetings scheduled with their Polish colleagues, as Table 4 shows. Some managers and key technical people attended many of these. As one commented, "We have meeting after meeting, then we eat lunch, and then we can start working." It was clear that the need to attend many scheduled meetings affected time spent on other tasks.

**Table 3**
Time in meetings per role.

|  | n | **Total hours spent in meetings** | Time in scheduled meetings | Time in unscheduled meetings and ad hoc conversations |
|---|---|---|---|---|
| Developer | 70 | **2.4** | 0.98 | 1.40 |
| Tester | 31 | **2.1** | 0.84 | 1.28 |
| Manager | 27 | **5.5** | 2.87 | 2.59 |
| Architects | 8 | **4.9** | 2.31 | 2.54 |
| Other | 6 | **3.7** | 1.98 | 1.75 |

**Table 4**
Daily group mode coordination

| Start | End | Team meeting |
|---|---|---|
| **08:50** | 09:00 | Operations |
| **09:00** | 09:15 | UX |
| **09:15** | 09:30 | Front-end |
| **09:30** | 09:45 | Back-end |
| **09:45** | 10:00 | Support meeting |
| **10:00** | 10:15 | Bug triage |

While the daily meeting was the most important coordination meeting, team members did not wait for the next daily meeting to address issues, as observed in a stand-up meeting:

"[UX developer], do you have anything to add?"

"No, I got answers on Slack yesterday."

While there was a need for frequent meetings, they also tried to implement architecture principles to limit the number of additional meetings. For example, the use of collaboration through APIs (application interfaces), was an idea intended to reduce the need for regular meetings across sites. One example was a Norwegian developer who maintained an API for a back-end part of the system that front-end developers in Poland could access. Although the idea was good in theory, in practice, the approach did not work as expected. The Norwegian developer explained, "We are supposed to collaborate through the API, but it is not enough—we need to talk. Therefore, we have weekly meetings with the other site."

*4.3 Availability of key people*

Based on the interviews, we found that a lack of access to key people due to distribution, time zone challenges, and key people being busy reduced the possibility for teams to call for meetings across sites. One Chinese tester explained that the testers in China wanted more meetings with the developers and managers in Norway than were happening at the time of the interview. Little time overlap (only 2-3 hours) between the working hours in Norway and China made this problematic. When interviewing the Norwegians, we found that the managers in Norway were involved in multiple projects with many meetings and were often required to travel, which made scheduling regular meetings with the Chinese challenging. One manager in Norway stated, "Testers N, NN, and NNN in China have asked to organize meetings. But I'm very busy and there is not much overlapping time during the day. So the meetings do not happen as often as they want." From the survey, we found that the managers, spent an average of 5 hours and 30 minutes per day in meetings, which represents a significant amount of their day. Managers in Norway not being available for meetings in critical periods frustrated the remote testers. In addition, because the managers indicated they were swamped with work, sometimes remote testers did not even ask for a meeting because they assumed the manager would be too busy to attend.

In some projects, when developers or architects on one site were too busy, they assigned a coordinator to handle communication with remote team members. The idea was that one person would handle and respond to all requests from the other site, and check with local staff as needed. This was perceived as challenging for the members having to communicate through this person. A Chinese tester explained:

"I cannot talk directly to the developers. I think this is a problem. I asked why, and they said that the development team does not want me to disrupt them, because they are just busy coding. So they just assign one coordinator to answer all of my questions. You can guess how difficult that is for me."

Projects being spread over multiple time zones was another challenge. In one of the projects, conducting a synchronous stand-up meeting was impossible. The team members were distributed between Norway,

China, and the US. The team tried three different techniques during the course of the study. First, they tried conducting one stand-up with participants from Norway and China, and then Norway and the US later in the day. Second, they tried a meeting with participants from Norway and China or Norway and the US, and then recorded the session so the other group could watch it later. Third, they used Slack, for which a Slack bot collected all members' statuses. The first technique was seen as too time-consuming, and the second technique was perceived as having little value for the partner watching the recording; the third, in combination with other meetings, was the most successful practice at the time of the study.

### 4.4 Unbalanced activity

#### 4.4.1 Unbalanced activity in meetings

Other phenomena that repeatedly appeared in interviews and observations were that some participants did not talk or contribute as much as others and that discussions happened mostly at the site that had the most people or key people. In some meetings early in the study, we observed that the Norwegians talked more and the Chinese talked less, even though we knew from the interviews that the Chinese had a lot of questions. When we interviewed the Norwegians, they told us that the Chinese team members did not raise issues in the meetings—issues that should have been raised. Building trust to enable remote team members to ask questions was seen as a challenge by team members at both sites. A test manager reflected on how to mitigate this challenge in virtual meetings: "In every meeting, I start the meeting by reminding the testers about the importance of asking questions." She applied this practice on a regular basis over a year, and later in the study, we were told that the Chinese team members were more active participants in the virtual meetings. Awareness of the importance of facilitating virtual meetings produced results.

Another facilitation practice in Geosoft was to rotate the facilitator role between sites. By rotating the facilitator role, Geosoft reduced the possibility of one site dominating the communication or decision-making. A manager from Norway posted on Slack,

"@PolishDeveloper Can you be the facilitator for today's retrospective meeting?"

#### 4.4.2 Unbalanced activity on Slack

When analyzing channels in the collected Slack logs from the product center distributed across Norway and Poland, we found that one-third (10 of 30) of the most active users had written 86% of the messages. Figure 4 shows user activity in the three essential channels: general, back-end, and front-end. Of the 10 most active project members, the two Norwegian tech leads were the most and third-most active users, and two Norwegian senior developers were the second- and fourth-most active. One Norwegian tech lead commented on her role and the need to coordinate with others:

"Some developers and testers are lacking the technical skills, so you need to help and support them."

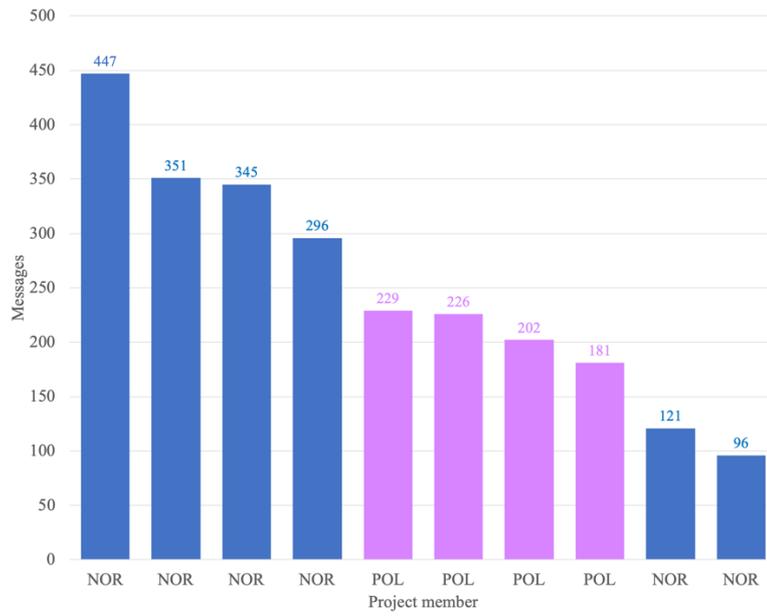

**Figure 4.** User activity across three channels. NOR= Norwegian, POL = Polish

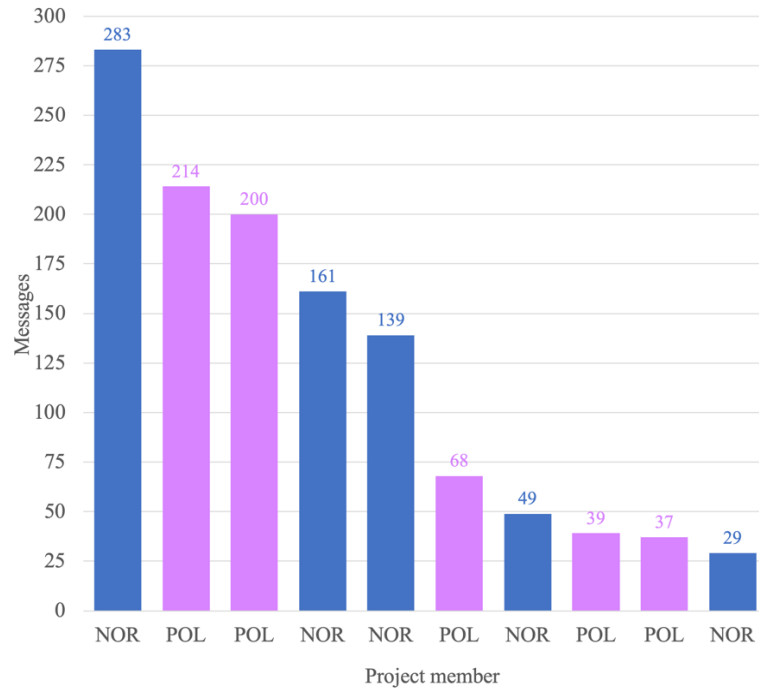

**Figure 5.** User activity in the Back-end channel. NOR= Norwegian, POL = Polish

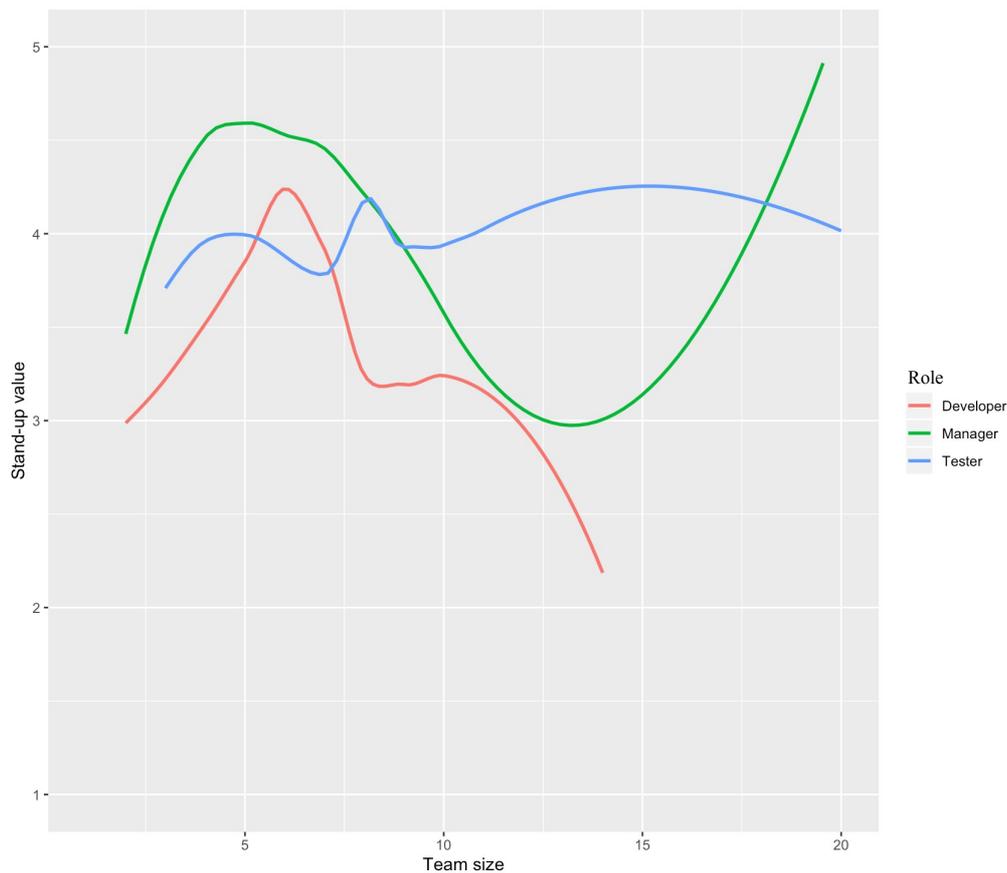

**Figure 6.** Visualization of team size and value of stand-up meetings for three roles

When analyzing each channel separately, we found that the back-end channel had more balanced user activity among the countries (Figure 5). One reason was that the Norwegian tech leads were posting in several channels and therefore, were the most active users across channels.

This imbalance was perceived by some team members as a problem and discussed in a group interview. A tech lead stated, "Some team members don't ask any questions [in the channels]; maybe it's a cultural difference."

The most active user was a Norwegian tech lead, whose main task became to support the developers—in particular, she spent a lot of time communicating with the Polish developers. An excerpt from the Slack logs gives an example of a Polish developer asking her for clarification:

"To me this seems like test data that has been added to the database by mistake... but maybe I'm wrong. @techLead?"

### 4.5 Perceived value of meetings

The number of participants in a meeting affected how valuable people perceived the meetings to be. For example, we observed that some retrospective meetings were much better when the number of people was not very high (5-7 people) and everyone contributed. With fewer people, we noticed that people paid more attention to what others were saying. One tester in China said, "I like our daily distributed meetings, because we only have three or four people. We can talk with each other, update each other on project status, and discuss important issues. For small teams, I think daily video meetings are the best. For large teams, maybe a weekly meeting or a Slack meeting is better.

Based on an analysis of team size and perceived value of daily stand-up meetings (from 1 - not valuable to 5 - very valuable), we found that developers, testers, and managers perceived the value of meetings differently and that team size matters. We show the result of the analysis in Figure 6.

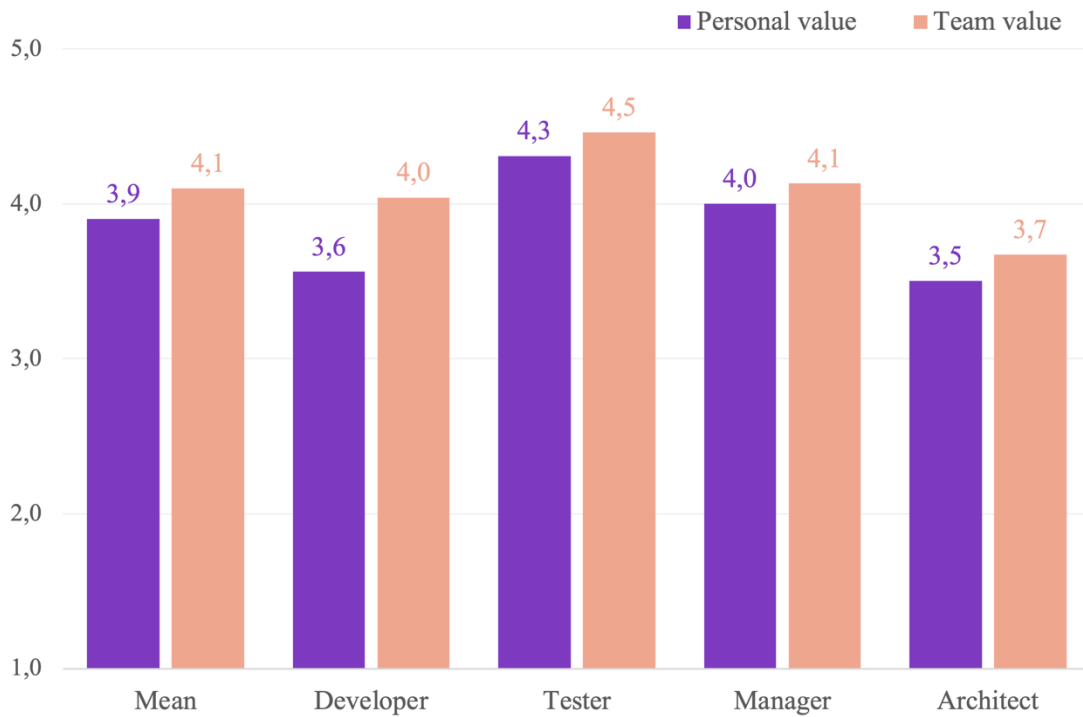

**Fig. 7.** Value of distributed daily stand-up meetings, personal value versus team value

For developers, those in teams with six members rated the daily stand-up meetings higher than for any other team size. They also perceived the value of the meetings to decrease when the team size was larger than 11. Interestingly, for testers, the daily stand-ups were seen as valuable independent of the size of the team; they rated these close to 4 for all team sizes. Managers rated the meetings as more valuable than developers for all team sizes.

To better understand the value of stand-up meetings we asked the participants to rate these meetings from both personal and team perspectives (see Figure 7). Participants of all roles rated their distributed meetings as being valuable (a value of 3 meant "neutral"). The results also showed that respondents perceived the distributed daily stand-up meetings to be more valuable for the team as a whole than for themselves individually. Moreover, testers were most satisfied with the distributed daily stand-ups.

We found it a bit surprising that the respondents rated their distributed stand-up meetings slightly higher in value than the local stand-up meetings. One explanation may be that, in the China–Norway case, the stand-up was the most frequent joint meeting between the sites, and was therefore the primary source of information on what was going on in the project. Also, in China, the testers conducted local stand-up meetings with the other testers. A Chinese tester explained how the information shared in the local stand-up was less relevant because the participants did not work on the same project. She explained, "The local stand-up meeting is most useful for the team leader, who gets to know what everybody is working on. But for me, it's not very helpful to know what they are doing. It doesn't really matter if I know or not."

*4.6 Co-location*

To overcome the barriers of distribution, all projects had a strategy for co-location. In the projects we investigated, team members co-located once or twice a year. In the case of China and Norway, the testers usually went to Norway and the managers to China. In the Poland–Norway project, the whole project co-located once a year, in addition to individual team members and managers traveling regularly to the other site. The distance and cost (direct cost and hours spent) of travel between Norway and Poland are much lower than between Norway and China. This cost particularly influenced the travel frequency, and was the main reason that there were more frequent trips to Poland.

In Geosoft, when distributed team members were co-located, they organized the most complex and challenging meetings during the co-location periods. Examples were discussing the roadmap, complex technology issues, and conducting a retrospective meeting. We observed retrospective meetings in Norway, Poland, and China. Because co-located retrospectives had fewer misunderstandings, helped build a higher level of trust, and had fewer technical issues, they also held more value. Even though the co-located retrospectives were seen as more efficient, they often lasted longer than when they were conducted as virtual meetings. Reasons may be that more issues were raised and the discussions on issues went deeper. When trying to understand why the virtual retrospective meeting was perceived as a challenging meeting, interviewees gave reasons. These included that the participants were not able to sit by the same whiteboard, were not able to use stickers, and were not able to have the same kind of informal discussions as in their co-located retrospectives. Several teams tried various tools for supporting the virtual process, but nothing seemed to help the situation.

We also observed meetings in which participants discussed product roadmaps and technical work that was groundbreaking or complex. The product roadmap described what new features would be offered to customers and when, and often initiated important discussions and clarifications about what the organization was trying to achieve. A typical question in a co-located roadmap discussion was *what input is needed for the roadmap (from the team's perspective)?* Complex topics discussed included security issues, how to reuse components, how to improve quality (technical debt), and documentation. Having everyone understand where the organization was heading was seen as a prerequisite for the virtual organization to function effectively; therefore, they invested a lot of effort in this when they co-located (e.g., dinners, trips, and other social events).

*4.7 Slack collaboration*

We found that team members often started to use new collaboration tools without explicitly discussing norms on how to use them. Motivation for using the tools and how the tool was used varied between roles. One manager said, "We use Slack to share knowledge, communicate frequently, and to enable continuous learning." The team members also used Slack to easily share files, especially screenshots. A Norwegian team member offering help to a Polish team member posted, "Yes, I can help. Do you have a screenshot of how this should look when it is done?" Some of the team members started to post information on Slack to have it accessible for future reference.

When it came to the use of Slack for coordination, a lack of agreement on how to use the tool led to frustration and confusion for some team members. In one co-located retrospective when the project members had used Slack for over a year, it became evident that it was necessary to formalize Slack guidelines. People had different understandings of how best to use the tool, and newly hired people had to get up to speed. Further, some found it difficult to know who should join which channel. In Table 5, we show the principles for use of Slack that the project members agreed on in the retrospective meeting.

**Table 5.**
Principles for the use of Slack agreed on by the project members in the Norway–Poland case

| Slack principle | Explanation |
| --- | --- |
| **Less communication of features and bugs in other tools.** | While they still used other tools (such as Microsoft VSTS) to discuss features and bugs, they wanted this type of conversation to occur only in Slack—for example, by creating new channels for discussions of specific features or bugs. They believed that it would be easier to browse the history of the discussions of an issue if the discussion was not spread across different tools. |
| **More open communication.** | Open communication in channels was favored over direct messages between two people. |
| **Each team should have a main channel.** | For example, front-end, back-end, and user experience team channels. |
| **More separate channels.** | Rather than posting too many messages in team channels or in the general channel, they wanted narrower and more specific channels. All channels should have a description to make it easy to know what issues to discuss in which channel. |
| **More short-lived channels.** | A developer who first started to work on a feature should make a new channel for that feature (and similarly, for bugs). All new feature- or bug-specific channels should be mentioned in the general channel. They decided that such short-lived channels should have a specific prefix, showing that they were discussions of features, for example. Each channel should be archived when the feature is implemented or the bug has been solved, to reduce the number of channels. |

The members agreed that there was a need to have specific channels—for example, more short-lived channels to discuss separate issues. At the same time, they did not want to have so many channels that it became overwhelming to deal with notifications or difficult to follow relevant discussions. Finding the best structure in Slack seemed important, but challenging. When investigating the current Slack structure, a manager explained, "A major challenge has been to find a balance in the number of channels and make the channels as relevant as possible." Another project member stated, "We have created channels for different topics so that team members do not have to explain the context every time they send a new message, but it does not work perfectly."

The members also discussed the importance of someone not involved in the discussion responding immediately when they knew the answer to a question rather than waiting for others to answer (for example, a person mentioned using @).

*4.8 Unscheduled meetings*

Unscheduled (spontaneous) meetings were one type of meeting that was seen as happening too infrequently in the distributed teams. We often observed such meetings in the co-located teams. Spontaneous meetings were the type that just happened because people met, for example, at the coffee machine, in the cantina, or because team members were sitting in the same area. In the interviews, we found that spontaneous meetings happened even more infrequently when a long time had elapsed since the previous face-to-face meeting. One Polish developer reflected on this: "When it has been a long time since a visit, it seems that [the Norwegians] forget we are here."

To mitigate the absence of co-location, some projects in Geosoft tried out a "virtual water cooler" by having a live video feed from each of the sites during working hours between two sites. The video cameras were located in places where people from each of the sites spent time during their breaks. While the practice was initially a success, the teams stopped using it after a while because it failed to facilitate discussions across sites, and the value of using the technology was low. Although the teams stopped using the video link, Slack seemed, somehow, to take over for facilitating such unscheduled and spontaneous meetings. In Slack, the conversation is always on during the working day, and seeing others in a channel made it easy to invite them to spontaneous meetings. As one developer posted on Slack at 8:39 a.m.:

"Ok, guys and girls. We will try to have a small session at 11 today to discuss architecture and layer responsibilities in the front-end code. Hope as many of you as possible are able to join. I will not create a formal meeting invite. The ones that are interested can call in using Skype. :slightly_smiling_face:"

*4.9 Awareness*

Slack enables some of the same awareness in distributed projects as in co-located projects with an open office landscape regarding what people are doing. One example is that team members used Slack to notify each other about their presence or absence. In 2016, a developer suggested, "Should we have an #out-of-office channel here on slack? Easier than sending email IMO." Such a channel was created two days later and team members actively used it to inform each other of their presence across sites. Table 6 shows an excerpt from a log. Team members also notified each other of why they were unable to attend team meetings (e.g., daily stand-up meetings).

**Table 6.**
Excerpt from Channel #Out-of-Office

| User | Time | Statement |
|---|---|---|
| 1 | 06:17 | I'll be bit late today. There is something called as 'Easter Bake Sale' at my daughter's school for her class today at 9:15 which I'm going to attend to. |
| 2 | 06:17 | Good luck with that:-) |
| 3 | 06:18 | Thanks!! |
| 4 | 07:35 | I'm running late today. I'll be in as soon as possible. |
| 5 | 08:08 | I'm down with fever. I'll try to work some from home when the drugs kick in. Available on slack and phone all day. |

By contrast, one team member who worked in a project not using Slack was frustrated by having to communicate via a busy manager, and frustrated by the extent to which that decreased awareness of what others were doing. She stated, **"We cannot directly contact the developers. So for these two projects, most of the time, we don't know what the developers are doing—we don't know what's going on."**

One important feature in Slack is direct messaging, and many participants used private messages instead of communicating in public channels. Because of data privacy regulations, we were not able to gain access to direct messaging logs to measure how many of the messages were personal-mode messages and how many were group-mode messages. However, one interviewee suggested that half of the messages on Slack were direct messages between two team members. Several others confirmed this estimate regarding the use of direct messages. The Polish and Norwegian sites offered opinions that opposed each other regarding the use of direct messages. Generally, we found that the Norwegian developers and testers wanted more of the messages to be in open channels, to increase the awareness of what people were doing and what discussions and problems that team members had. However, the Polish project members appreciated having private conversations. The Norwegian manager said, "One should use the 'public' channels and not direct messages, to increase learning."

The managers and tech leads hoped that with more messages in open channels, others could better understand what was going on, could learn from the discussions, and that team members would need less support over time. However, changing an established practice was hard, as one tech lead explained when talking about his colleagues in Poland:

"I try to encourage them to write messages in open channels, but they still continue to send personal messages."

One explanation for why the Norwegians wanted the discussions to be visible to everyone may be that they had used agile methods and social software for a longer period and, therefore, had experienced the benefits of open team communication, while the developers in Poland were accustomed to more one-on-one communication in their teams and a more hierarchical organization. Another explanation may be that remote developers in Poland noticed that the Norwegians answered more quickly when contacted directly, as one Norwegian developer suggested:

"While I try to answer as quickly as possible in the channels, I give faster feedback when a person sends me a direct message."

## 5. DISCUSSION

We have described group-mode coordination mechanisms in GSE by conducting research on the use of meetings and the tool Slack in a global company located in Europe, Asia, and the United States. Further, we have conducted a detailed investigation of the collaboration between Norway and Poland (nearshore) and between China and Norway (far shore).

We surveyed how much time the participants spend in meetings and observed different types of meetings. Developers, testers, and architects described spending more time in unscheduled meetings than in scheduled ones, which is beneficial for solving more complex tasks [95]. Further, according to Eisenbart et al. [21], discussions in unscheduled meetings tend to be more focused and to lead to more effective decision-making than discussions in scheduled ones, which makes unplanned coordination valuable in global teams.

We found that the employees of Geosoft spend on average 7 hours and 45 minutes per week (1.55 hours per day) in scheduled meetings. Not many studies report on the hours spent in meetings. Rogelberg et al. [70] found in a 2006 study across various sectors (both private and public) that employees spend on average 5 hours and 36 minutes per week in scheduled meetings. This suggests that GSE employees spend around 38% more time in meetings than employees in other companies, which makes sense because of the complex tasks and need for coordination in global software projects. As the time spent on programming and meetings by the developers in our study is similar to the results of another study of programmers across many companies [90], it is reasonable to suggest that Geosoft is a representative company when it comes to a how a typical workday is organized.

The managers reported spending 27 hours and 30 minutes per week in meetings (5.5 hours per day), which suggests that meeting attendance for managers in global software companies represents an enormous proportion of their time at work. Dahl and Lewis [13] found in 1975 that managers spent 59% of their time in meetings. In a study of six CEOs in 1973, Mintzberg [53] found that managers spend up to 70% of their time in meetings. McCall et al. [50], as cited in [78], reported that "unscheduled meetings, or informal contacts, represent the largest time-consuming activity at middle to lower management levels." People spending a lot of time in meetings caused some of the coordination challenges we observed in Geosoft, and we discuss that issue later.

Agile software development relies on frequent interactions and mutual adjustment, and since distance makes people communicate less [61], virtual teams need tools that can mitigate distance and lower communication barriers—i.e., link different parts of the organization [95]. In Geosoft, we found Slack to be one of the most important collaboration tools. Further, we found that projects co-located once or twice a year and that they organized the most complex and challenging meetings during the co-location periods. Calefato et al. [7] also argued that face-to-face meetings are essential for having more in-depth discussions. Meeting face to face is important because distributed team members build relationships when they meet, and many distributed software development teams try to balance agile and distributed approaches by having regular visits to improve coordination [83]. Matthiesen et al. [49] studied a GSE setup between India and Denmark and found that whether interruptions on IM tools were perceived as normal or as negative disruptions depended on the quality of the relationships between the distributed colleagues.

We found that the agile teams using Slack could rely on mutual adjustment, which is the core coordination mechanism in agile software development [59]. Most often, we found the communication on Slack to be unscheduled, as communication was frequently triggered by someone asking a question, sharing information, or participating in a discussion. This way, Slack partly mitigates the problem with the lack of unscheduled meetings in GSE. While unscheduled conversations dominated in Slack, team members also scheduled Slack conversations or Slack video calls at specific times. Additionally, Slack supports an impersonal mode of coordination through, for example, automatic messages posted on Slack by bots and integration with other systems (e.g., the team members were notified when database test exports failed). Slack accordingly supports all the coordination modes suggested in the framework defined by Van De Ven [95] (see Figure 1). We found Slack to support coordination in the distributed projects and therefore argue that such collaboration tools serve to enable agile methods in a distributed context. Our research shows that some users were very active, while others posted very few messages. We found that language skills and knowledge level in particular influenced how active people were on Slack. Further, experienced team members favored messages in open channels (group mode), while less experienced people favored more direct messages (one-to-one communication, i.e., personal mode). Our data suggests that openness and transparency are building blocks of collaboration and trust in distributed agile projects.

We now discuss the case in light of our research question: *What are the group-mode coordination challenges in global software development projects?*

We will discuss challenges for unscheduled and scheduled meetings first and follow with challenges associated with using Slack.

### 5.1 Challenges of unscheduled and scheduled meetings in GSE

We found that new meetings emerged and changed over the course of the longitudinal study. That scheduled and unscheduled meetings change over time is confirmed by Moe et al. [56], who found that new and changed meetings emerge both from the top down and from the bottom up. Our findings are also consistent with those of Jarzabkowski et al. [35], who argued that coordination mechanisms do not arise as ready-to-use procedures but are constituted as actors go about the process of coordinating. Therefore, a retrospective meeting is one key meeting that functions as an arena for changing and improving group-mode coordination. Based on our findings, we identified three main challenges to scheduled and unscheduled meetings in GSE:

- Low availability of key people in far-shore projects
- Meeting facilitation is missing in virtual meetings
- Absence of organizational support for unscheduled meetings

#### 5.1.1 Low availability of key people in far shore projects

To organize effective meetings, the right people must be able to attend. In the far shore case, we found time zone issues and that key people and experts at the remote site were too busy because they spent many hours in meetings, which resulted in other meetings being declined or postponed. In one project, the participants had no overlapping working hours because they were distributed between China, Norway, and the United States. Consequently, it was not possible to organize joint virtual meetings. Some reasons for experts or key people being too busy were that they were involved in several projects and therefore attended many meetings and that their role as managers gave them little flexibility for organizing or attending new meetings.

When teams in the far shore case did not get access to remote key people, it seemed that they also interacted less with other teams' external resources, compared to what we observed in the nearshore case. That is one reason why introducing coordinators between the sites was a bad idea that made the situation worse for the remote team members. In the nearshore case, it seemed that everyone was able to get in contact with everyone else. Our findings are consonant with that of Smite et al. [82], which is that many meetings and forums increase the amount and frequency of communication between teams outside of meetings.

While many small issues and questions can be solved by e-mail and other collaboration tools, more complex problems cannot. Van de Ven et al. [95] highlighted the importance of meetings, in which complex problems are solved: when task uncertainty increases, scheduled and unscheduled meetings should serve as the main coordination mechanism. Due to the challenge of conducting meetings, problems take more time to solve, and subsequently the speed of development is reduced in far-shore projects. Further, when teams get "out of sync," they are likely to experience problems with the coordination process and to fall below their expected productivity level [47]. Other research in software development has also found that the absence of experts leads to miscommunication and delays in coordination with the offshore site [82].

#### 5.1.2 Meeting facilitation is missing in virtual meetings

We observed that many participants regarded poorly conducted meetings as less valuable. Indications of poor meetings were participants being silent during a whole meeting (unbalanced contribution), people not paying attention, or meetings not concluding in a timely manner. In the survey, every role rated the value of the meeting as higher for the team as a whole than for the individual (Figure 7), meaning that participants believed the meeting to be valuable for the team but less valuable for themselves. One explanation is the lack of a good meeting process and meeting facilitation, which previous studies on such meetings have confirmed [92]. We found that the larger the meetings, the less value they gave. Paasivaara et al. [66] also found that GSE meetings involving too many participants with disjointed interests and concerns function poorly.

Many researchers have studied the meeting process. For example, a recent study of software projects found that planned meetings were associated with a negative perception of productivity by developers [52]. Further, Wittenbaum et al. [99] concluded that "facilitating the successful coordination of group members may be the key ingredient to improving group performance." Sauer and Kauffeld [76] discussed the role of a meeting facilitator, arguing for the importance of a decentralized interaction structure in team meetings and asserting that facilitators must make sure that everyone speaks. Coaching to change virtual meetings can, therefore, yield substantial and enduring improvements in team effectiveness. We observed a few attempts of participants to facilitate a meeting by encouraging others to speak up or by rotating the facilitator role in a workshop. One manager who made a practice of reminding the Chinese team members to ask questions achieved an improved situation over time. Coaching interventions must be implemented when the team is ready [29]. As such, a retrospective could be a good time for meeting interventions. However, challenges with meeting facilitation were not discussed in retrospectives or mentioned in the interviews.

Even though we did not find much data on attempts to improve meetings, it was obvious that managers were aware of the limitations of discussing complex matters in virtual meetings. As a consequence, many problem-solving meetings were scheduled during regular face-to-face visits. We found one strategy for reducing the number of meetings that consisted of using an API architectural strategy, but this did not seem to work.

### 5.1.3 Absence of organizational support for unscheduled meetings

The importance of unscheduled meetings is consonant with Van de Ven et al. [95], who found that teams must rely on unscheduled meetings to a greater extent than on scheduled meetings when task uncertainty is high. Group-mode coordination by unscheduled meetings is ensured by team members and teams sitting together in the same office [62, 93]. However, in GSE, that is not possible. We found attempts to simulate virtual presence, such as the virtual water cooler, but the practice stopped, as it did not yield the expected effect. Slack seemed to facilitate virtual presence because it was easy to communicate what was going on or whether someone was out of office and even to conduct meetings. Slack is further discussed in the next chapter. We found that co-location for a period helped create awareness of who was doing what and seemed to initiate more unscheduled meetings in the period after a co-location event.

One limitation of unscheduled meetings was that developers and testers were sitting in an open work area. If a person received a request for an unscheduled meeting, the individual needed to find an available meeting room. When some people wanted a video room, they usually had to reserve the room in advance. Our findings are consonant with Smite et al. [82], who studied multiple teams in a distributed setting and found that it was challenging to have spontaneous meetings because the video conference rooms were frequently fully booked. Further, because the organization they studied had a culture in which managers and key people attended many scheduled meetings, this left no time for unscheduled activities. Smite et al.'s [82] study at Ericsson confirmed our finding, which was that priority was given to scheduled meetings.

### 5.2 Challenges with the use of Slack in GSE

We found that Slack supports problem-focused communication. This type of communication involves problem-solving discussions in which team members discuss knowledge and solutions and is linked to positive team outcomes [36]. Further, Slack supports all coordination modes (impersonal, personal, and group mode) in the distributed project. However, while there are many advantages of using Slack, we found some barriers to using the collaboration tool. Based on our findings, we identified three main challenges associated with the use of Slack in GSE:

1) Language
2) Unbalanced activity
3) Lack of formalized coordination procedures

### 5.2.1 Language

Language is a challenge often reported in distributed projects [43, 61]. This was also evident in our study. When team members in distributed projects have to communicate in a second language, the quality of communication declines [60], which explains why some messages in our study were in Norwegian or Polish. Team members also used their native language in Slack, but this behavior excluded other team members from the conversation and weakened the ability to understand what was happening on another site. When one site does not know why there is little progress from members at the other site, the level of trust is reduced [60]. Further, we found that since the Norwegians had used English as a working language for a long time, many were proficient in English, which may explain why they dominated the conversations more in the Slack channels. Our findings square with a study of a distributed agile project [88] that found that people who were confident in a second language dominated more in meetings. Our findings suggest that English-language skills and unbalanced activity on Slack were related. We now continue on to discuss the unbalanced activity.

*5.2.2 Unbalanced activity*

In addition to language barriers, another reason for the imbalance between the two sites was that the tech leads and senior developers were in Norway while more junior developers were located in Poland. Our findings are in agreement with those of Smite et al. [82]: new hires and less experienced people communicate less frequently than more experienced team members. Apart from the difference in expertise, the Norwegians had worked together for longer and so probably had stronger ties as well. Strong ties and higher knowledge levels influence the frequency of communication [25], which probably resulted in more frequent communication among the Norwegians.

While analysis of the Slack logs showed an imbalance, we believe the communication was more balanced in reality. The Polish developers' and testers' preference for direct messaging (personal mode) over messages in channels (group mode) affected the total analysis; because the direct messages could not be included, there were fewer messages from the Polish team members in the public channels. Being less experienced and therefore having a greater need for help from an experienced person may also be an explanation for these team members' appreciation of direct messaging; that is, the mode allowed them to receive feedback more quickly. The activity imbalance in Slack between the countries can also be explained by other relevant streams of research (e.g., diffusion-of-innovation literature) that address the adoption of methods and technology [72], as the Norwegian team members had used social software for a longer time.

When solving complex tasks with a high degree of uncertainty, as in a distributed software project, the team must rely on a high level of mutual adjustment in both personal and group modes [95]. We found a high but unbalanced use of both modes, because the Polish developers and testers appreciated the vertical personal mode over the group mode. While the Polish team members were encouraged to reduce their use of the personal mode and increase their use of the group mode with respect to Slack, they showed resistance because the personal mode was more comfortable for them and yielded faster feedback.

*5.2.3 Lack of formalized coordination procedures*

The team members under study did not specifically discuss how to use Slack when they adopted the collaboration tool. There were few discussions on the use of Slack in team and management meetings Formalized procedures on how to collaborate were not created until the team members had been using the tool for over a year. Such procedures are important, and relying on the right communication norms to emerge by themselves is not wise, as the team members might be unaware of the norms or have a different understanding of them based on their cultural background [87, 94]. Different norms are likely to affect how people coordinate [100]. A different understanding of using direct versus group messages is an example. One example of Slack principles that a team agreed on in a retrospective meeting can be found in Table 5.

New team members might be unfamiliar with Slack and therefore continue to use tools to which they are accustomed (e.g., e-mail). For teamwork in distributed teams to be successful, all team members should be involved in discussions on a dedicated tool. As many companies are now starting to implement BizDev teams [23] (which means that business people are becoming a part of teams), it is especially important that team members from the business side use the same collaboration tool.

### 5.3 Implications for Practice

#### 5.3.1 Meetings in distributed teams

When meetings were conducted efficiently, they tended to enhance the team spirit. Finding the time to conduct daily stand-up meetings in distributed teams, whose members are spread over multiple time zones, is challenging. For nearshore teams, daily stand-up meetings are easier to hold, but we saw in our research that these meetings were often centered on sharing statuses from the different sites. It is more valuable to share such status information using Slack, so that meetings can then be used to ask questions, solve problems, and find solutions together.

If the facilitator of meetings is always at one site, this may reduce the meetings' value for the other participating sites; it is therefore valuable to have the facilitator rotate between the different sites. Bjørn et al. [5] also found it was essential to regularly rotate mediators or boundary spanners to make collaboration in GSE work across sites. In team meetings, special attention should be given to facilitating conversation so that everyone is encouraged to speak. In a study on successful teamwork, Google found that to make a team productive, every member should over time speak for roughly the same amount of time [20].

We found that distributed teams were significantly larger than co-located teams, which may have created a challenge in holding team meetings, as excessive numbers of participants reduced the value of the meetings. We found the team size for distributed teams to be on average 8.7 members; it is fortunate this number was not higher, because research on software teams has found that teams with more than 9 members are less productive [69]. Global projects should strive not to have teams that are too large. One task for the facilitator should be to make sure that only relevant people are attending the meeting and that all attendees really need to be there.

By understanding how much time agile team members and managers spend in scheduled and unscheduled meetings, it is possible to better understand how much time is used for coordination. The strategy of introducing coordinators (a role that is supposed to coordinate the communication between remote team members and local experts) does not help the situation but rather makes people frustrated. To be able to solve this challenge, there are two possible solutions: key people need to be more available for the remote teams or the remote teams need to be given more authority and responsibility to become more autonomous.

#### 5.3.2 Slack supports agile teams

A systematic literature review reported that IM tools such as Slack [27] accelerate communication and cultivate rapid feedback in global teams. Our findings also suggest that Slack supports building stronger and more autonomous teams, which is a prerequisite for the success of agile teams. Slack supported the teams and their teamwork by

- increasing team awareness and transparency by supporting constant information sharing;
- increasing the speed of feedback;
- facilitating network building (both team internal and external networks);
- increasing the awareness of who knows what, which is essential for high-performing teams [41]; and
- reducing the need for e-mail and other channels.

Team awareness consists of an understanding of the activities of others [19] and is the result of recurrent processes of information sharing within a team [74]. In our research, the ongoing information sharing in Slack helped build team awareness and increased transparency regarding what was going on. While activity in Slack channels supported strengthened awareness of what was going on in the team, which is important for coordination by feedback, we also found one channel dedicated to sharing information on non-job activities. Distributed team members who have met and know each other personally have stronger ties and communicate better [18, 25], and Slack supported this finding; for example, when team members talked about their family members in the "out of office" channel (e.g., being sick, having a bake sale), the personal ties of the virtual members became stronger. Dittrich and Giuffrida also found that social discussions with IM tools helped build good relationships between distributed team members [17]. Further, when team members at a remote site know why a person is absent, their level of trust is maintained. Awareness of what is happening and who is doing what also seemed in our research to initiate unscheduled meetings, which we found was more challenging to conduct when teams were distributed. Our findings suggest that Slack facilitates constant informal communication through formal channels, which improves communication in agile distributed projects [67].

An essential tenet of agile methods is transparency. Slack facilitates the group mode of coordination because all of the virtual team members can engage in discussion. By contrast, team members not included in an e-mail thread, for example, may miss out on relevant information, or they may inadvertently withhold valuable information that they do not know other project members need because they are unaware of ongoing discussion. The team members received fast responses to their questions on Slack. Frequent communication builds trust and awareness of tasks and how they affect each other [51]. Further, project participants kept each other informed on what they were doing, which increased the transparency between sites. Increased transparency also builds trust, which is vital for success in distributed teams [60].

The team members experienced the introduction of Slack as a positive change; however, at the same time, it was also an opportunity for senior members and tech leads to assume more rigid control over each team member. One team lead reported asking many questions to ensure that everything was understood and agreed upon. Our findings are in agreement with those of Moe [54] and Barker [3], who pointed out that self-managing teams may end up controlling group members more rigidly than teams that rely on traditional management styles. When team members in virtual projects believe they are being controlled, they lose trust in their counterparts [60].

Finally, an implication of this study is that for companies to increase the likelihood of successful Slack adoption, managers in GSE projects need to support the introduction of such tools, e.g., by actively facilitating a bottom-up process of creating Slack guidelines and removing barriers to using Slack. Furthermore, managers may analyze the usage trends in Slack, such as by examining the number of messages sent and received by the teams during a month and observing trends in the number of messages at different times during working hours. Such analysis also makes it possible for managers to identify people who send too few messages and to help them master the tool.

*5.4 Limitations*

Our research examines "teams in the wild"—or teams embedded in large organizations and broader sociotechnical systems [73]—which are important for understanding how agile team members interact in GSE. As in all empirical studies, though, we need to consider some limitations. A first limitation is that we used a single-case design. Therefore, the general criticisms of single-case studies, such as uniqueness and special access to key informants, may also apply to our study. However, the rationale for choosing this company and conducting a detailed study of four collaborating sites (Poland–Norway and China–Norway) was that the company and its projects represented a critical case for exploring the group-mode coordination challenges in global software development projects. The Poland–Norway and China–Norway sites developed products based on different technologies and customer relationships. Because the same phenomena were reported across many teams, it is likely that other medium-sized companies and projects will experience the reported phenomena and that the conclusions in this study will prove useful.

Another possible limitation is that much of the data collection and analysis was based on semi-structured interviews and participant observation. The consequence of this limitation is that the results have been influenced by our interpretation of the phenomena observed and investigated. The authors attempted to mitigate bias by using an interview protocol for each interview (Appendix A). We strived to ask questions in a friendly, nonthreatening way, focusing on posing questions such as "how" and not "why," which helped create a rich dialogue and reduce interviewer bias [4]. For observations, biases can occur to the point that a researcher makes biased conclusions based on inadequate data and prior subjective opinions, or the observed may act differently with an observer present, which is known as the observer effect [68]. To reduce observation bias, Becker [4] suggested noting down all relevant interaction so that the researcher might be able to distinctively remember details. While observing the project members, one of the researchers noted down as much as possible, and we also used an observational protocol when we observed meetings (Appendix B).

Further, the use of multiple data sources made it possible to confirm evidence for episodes and phenomena. The study included observing, talking to, and interviewing team members and managers from several sites and teams, which made it possible to investigate phenomena from different viewpoints as they emerged and changed, thus reducing this limitation. There is a risk that our findings could also be explained by factors that evaded our attention. However, giving feedback to the observed teams and discussing our interpretation of what was going on helped validate our conclusions.

# 6. CONCLUSION AND FUTURE WORK

A key aspect of successful GSE is coordination. In this paper, we shed light on coordination in global teams by presenting a longitudinal study of meetings and collaboration tools in a large company with software development sites in China, Europe and the United States. In sum, our article offers the following contributions. First, we provide readers with some background on teams in agile GSE and on how much time project members spend in meetings. Second, we discuss challenges associated with conducting scheduled and unscheduled meetings. Third, we explain how the collaboration tool Slack is used to coordinate in GSE and associated challenges. We illustrate our explanations with data from surveys, observations, interviews, and chat logs.

Meetings and ad hoc conversations provided an important venue for coordination in the GSE projects and teams we investigated. The average team size was seven members, and distributed teams were significantly larger than co-located ones. Our results show that the project members spent 7 hours and 45 minutes in scheduled meetings and 8 hours and 54 minutes in unscheduled meetings and ad hoc conversations during a typical work week. Managers spent as much as 14 hours and 21 minutes in scheduled meetings and 12 hours and 42 meetings in unscheduled meetings per week.

We conclude that organizing scheduled and unscheduled meetings is challenging in a far-shore context because of limited access to remote experts and key people as well as technical and organizational barriers to conducting meetings spontaneously. Further, scheduled meetings need facilitation to reduce imbalance between participants and to increase the meeting value. However, Slack reduced some of the barriers to inviting members to unscheduled meetings by showing who was online; team members also used the tool to inform each other of their presence and what they were working on. Waiting for others because of time zone barriers is a known challenge in GSE. We argue that unscheduled meetings are the most important in complex GSE projects because they increase the speed of decision-making and information sharing. Surprisingly, we found the daily stand-up meeting, which was both held locally and distributed, to yield more value in the distributed context, which indicates that who is in the meeting and what is discussed are more important than whether the meeting is distributed or not.

In our case study, the team members continuously communicated and coordinated with other virtual team members. We conclude that it is essential for the team members to be comfortable with the tools they are using and that there should be formalized procedures for the use of such tools so that everyone can benefit from them. According to our results, Slack supports frequent communication and fast responses within and between teams and their stakeholders, which in turn benefit GSE companies.

This study highlights challenges associated with scheduled and unscheduled meetings in a nearshore case (Poland–Norway) and a far-shore case (Norway–China). Our findings suggest that even in mature agile GSE companies using new tools and coordinating with both scheduled and unscheduled meetings, the same old barriers—such as language, unbalanced activity, and difficulty with facilitating communication— still appear. It is interesting that these problems remain evident, and future research should target how to mitigate such never-ending hurdles. Further, future work should analyze group interactions in meetings, by such means as using existing coding schemes for verbal behavior in project team meetings [39]. Researchers should also study how teams in global projects construct norms and identities in meetings.

Another potentially relevant concept to pursue in future work would be to better understand the role of team members' direct messages when using tools such as Slack, because direct messaging increases speed but reduces transparency. How the use of personal and group modes of coordination influence team performance are highly relevant for researchers to investigate. There is also a need to compare the numbers of hours that participants report spending on coordination with observations of how much time they actually spend coordinating, as an earlier study suggested that people underestimate their time when self-reporting [20]. Finally, while both meetings and the use of Slack are valuable coordination mechanisms, both are potential interrupters of flow and may be perceived as disruptions and cause fragmented work [49]. The balance of coordinating activities (such as responding to other team members' questions and attending meetings) versus immersion in one's own tasks is a difficult and relevant topic for future research.

ACKNOWLEDGMENTS

We are grateful to all the participants in the case study. We also thank Mehdi Noroozi for help with preparation and initial analysis of the Slack logs. This work was supported by the projects A-teams and Digital Class and the Research Council of Norway through grants 267704 and 309631.

# APPENDIX A

**Interview guide**

| Part | Question |
| --- | --- |
| 1. Introduction | Present ourselves. |
|  | Say thank you for participating. |
|  | Assure confidentiality. |
|  | Ask permission to tape record. |
| 2. Warm-up | How long have you been on this project? |
| | What is your role in the team/What do you do? |
| | Who do you see as your team members? |
| | Do you collaborate with other teams? |
| 3. Coordination | How do you communicate with people at other sites? (coordinating mechanisms / Van de Ven, rules for QA, instant messaging, informal ad hoc conversations etc.) |
| | How do you know if the developers are available for questions? (awareness) |
| | Who do you contact? How? |
| | How difficult is it to get the information you need from people at other sites? |
| | Benefits and challenges with instant messaging? |
| | Meetings vs instant messaging? |
| | How important is it to meet team members face-to-face? (How often?) |
| | Which meetings do you have together (with the other site)?<br><br>• What is working well, what could be improved?<br>• Show and tell? |
| | Tell me about your local daily meetings<br><br>• What is working?<br>• What is not working?<br>• What have you done to improve them? |
| | Tell me about your distributed daily meetings<br><br>• What is working?<br>• What is not working?<br>• What have you done to improve them? |
| | Tell me about your retrospective meetings. |
| | Tell me about the planning meetings. |
| | What are some challenges of working distributed? |
| | How are tasks allocated? |
| | When you receive tasks to do, how is it communicated to you?<br><br>• By who? |
| | After you receive a piece of work, do you need to collaborate with others to plan the work? |
| | When you have completed a piece of work, how do you report it? |
| | How do you get feedback on your work? When? |
| | Can it happen that you need to divide the work among other colleagues?<br><br>• If so, how do you divide the work, with whom do you share it, and how do you coordinate it? |
| | When you have completed a piece of work, how do you report it?<br>Can it happen that more people gets involved in your work during the time you are working on it?<br>• If so, how and who? |
| | How do you solve problems? |
| | What do you think of the information flow in the project? |

| | Do you have an overview of what others are doing? |
|---|---|
| | What can you think of that could improve the effectiveness of the teamwork or the project in general? |
| | How does the team make decisions? |
| | Do team members show interest in other individuals' tasks? How? |
| Closing | Do you have any questions for me? |
| | Is there anything else you would like to discuss that was not covered by the questions asked? |

# APPENDIX B
## Observational Protocol

| Topic | Question |
|---|---|
| Space | What is the layout of the physical room? |
| | How are the actors positioned? |
| Participants | What are the names and relevant details of the people involved? |
| | Is someone acting as a leader or facilitator? |
| Activities | What are the various activities? |
| | What are the discussions? |
| Objects | Which physical elements are used? |
| Acts | Are there any specific individual actions? |
| | What are the ways in which all actors interact and behave toward each other? |
| Events | Are there any particular occasions or anything unexpected? |
| Time | When does the meeting start? |
| | What is the sequence of events? |
| | When does the meeting end? |
| Goals | What are the actors attempting to accomplish? |
| Feelings | What are the emotions in the particular contexts? |
| | How is the atmosphere? |
| Closing | How is the meeting ended? |
| | Is there a post meeting? |

# APPENDIX C
## A selection of the survey questions

- What is your location?
- Do you work in a team?
- What do you consider to be your main role?
- What development method do you use?

- Consider the team you spend the most time in. Is your team...

co-located?

distributed across sites?

• How many meetings do you attend during a regular work-day (including daily stand-up meetings)?

Do you attend stand-up meetings (aka Daily Scrums?)
- Yes, local stand-up meetings
- Yes, distributed stand-up meetings
- Yes, local and distributed stand-up meetings
- No

How valuable do you think the **local** daily stand-up meetings are? (From 1 - Not valuable to 5 - Very valuable)
- How valuable do you think the meeting is for you personally?
- How valuable do you think the meeting is for the team as a whole?

How valuable do you think the **distributed** daily stand-up meetings are? (From 1 - Not valuable to 5 - Very valuable)
- How valuable do you think the meeting is for you personally?
- How valuable do you think the meeting is for the team as a whole?

Roughly how many hours do you spend on the following activities during a typical work day?
- Time in scheduled meetings (including daily stand-up meetings)
- Time in spontaneous unscheduled meetings and ad hoc conversations
- Time doing testing
- Time doing programming